\documentclass[aps,prb,reprint,superscriptaddress,showpacs,citeautoscript]{revtex4-2}

\usepackage[dvipdfmx]{graphicx}
\usepackage{color}

\usepackage{amsmath}
\usepackage{accents}

\makeatletter
\def\widebar{\accentset{{\cc@style\underline{\mskip10mu}}}}
\makeatother

\begin{document}

\title{Impact of in-plane disorders on the thermal conductivity of AgCrSe$_2$}


\author{Shota~Izumi}
\affiliation{Department of Materials Science, Osaka Metropolitan University, Osaka 599-8531, Japan}

\author{Yui~Ishii}
\email{yishii@mat.shimane-u.ac.jp}
\affiliation{Co-Creation Institute for Advanced Materials, Shimane University, Shimane 690-8504, Japan}

\author{Jinfeng~Zhu}
\affiliation{School of Physics and Astronomy, Shanghai Jiao Tong University, Shanghai 200240, China}

\author{Tsunemasa~Sakamoto}
\affiliation{Department of Materials Science, Osaka Metropolitan University, Osaka 599-8531, Japan}

\author{Shintaro~Kobayashi}
\affiliation{Japan Synchrotron Radiation Research Institute (JASRI), Hyogo 679-5198, Japan}

\author{Shogo~Kawaguchi}
\affiliation{Japan Synchrotron Radiation Research Institute (JASRI), Hyogo 679-5198, Japan}

\author{Jie~Ma}
\affiliation{School of Physics and Astronomy, Shanghai Jiao Tong University, Shanghai 200240, China}

\author{Shigeo~Mori}
\affiliation{Department of Materials Science, Osaka Metropolitan University, Osaka 599-8531, Japan}



\begin{abstract}
	Superionic conductors have recently attracted renewed attention for their use as thermoelectric materials due to their extremely low lattice thermal conductivity.
	Of central interest is why the superionic conductors exhibit such low thermal conductivity, and competing mechanisms have been proposed thus far.
	In this study, we investigate the effects of Cu and Au substitution for Ag site on the crystal structure and thermal properties of AgCrSe$_2$, which exhibits superionic conduction of Ag ions.
	We show that Au substitution significantly reduces the lattice thermal conductivity of AgCrSe$_2$.
	Powder structure analysis using synchrotron x-ray diffraction reveals that Au substitution increases the anisotropic atomic displacement parameter of Ag ions along the $a$ and $b$ axes. 
	This result indicates that the amplitude of in-plane vibrations is enhanced, which is attributed to increased anharmonicity in the potential energy around Ag ions.
	The enhanced vibrational amplitude also suggests a reduction in the force constants between Ag ions.
	Consequently, the enhanced anharmonicity not only shortens the phonon lifetime ($\tau$) by increasing phonon-phonon scattering, but also increases the  number of low-energy phonons, which further contributes to the reduction of $\tau$.
	This anharmonicity mechanism is applicable to other superionic conductors exhibiting ultra-low thermal conductivity, promoting their widespread use as thermoelectric materials. 
\end{abstract}


\maketitle

\section{Introduction}
	Materials with superionic conductivity often exhibit extremely low thermal conductivity ($\kappa$) and are therefore expected to be used as high-performance thermoelectric materials.
	For example, Cu$_2$Se, $M$CrSe$_2$ ($M$ = Ag and Cu), and Ag$_8$SnSe$_6$ are the superionic conductors that have been known to exhibit extremely low $\kappa$ of 0.1$\sim$1 W K$^{-1}$ m$^{-1}$ at 300 K \cite{Cu2Se,superionic1_AgCrSe2,superionic2_AgCrSe2,AgSnSe}.
	In order to promote the widespread use of thermoelectric conversion materials, it is desirable to clarify the mechanism behind the low $\kappa$ of these materials, and several competing mechanisms have been proposed thus far \cite{Cu2Se,anmarmonicity_prl,INS-Ag-PNAS,AgSnSe,Xie,Mg3Bi2,kappa4,AgCrSe2_computation}.

	One explanation is the liquid-like ionic diffusion that directly scatters phonons, which has been first mentioned in Cu$_2$Se\cite{Cu2Se}.
	On the other hand, several reports have proposed lattice anharmonicity mechanisms \cite{anmarmonicity_prl,INS-Ag-PNAS,AgSnSe,Xie}.
	For example, in Ag$_8$SnSe$_6$, where Se ions are responsible for ionic conduction, diffusive properties in the lattice dynamics have recently been reported, and anharmonic vibration of the Se ions has been pointed out to be essential for realizing the extremely low lattice thermal conductivity \cite{AgSnSe}.
	The importance of the lattice anharmonicity has also been emphasized in Mg$_3$Bi$_2$ \cite{Mg3Bi2}.
	Localized vibrations of the conducting ions in quasi-2D potential wells have also been proposed \cite{kappa4}.
	However, these proposed mechanisms have been highly controversial, and the nature of anharmonicity as well as its microscopic mechanism on phonon scattering remain unclear.	
	It is indispensable to focus on a system that allows systematic tuning of the chemical composition and to clarify how the conducting ions affect the thermal conductivity.

	In general, the physical properties associated with phonons are often obscured at high temperatures due to the large number of thermally populated phonons, many of which are not directly related to the properties of interest.
	Particularly, the majority of lattice thermal conductivity is owed to long-wavelength acoustic phonons.
	Therefore, it is crucial to know the low-temperature properties to discuss the thermal transport by phonons.
	Nevertheless, most studies on thermoelectric  materials with superionic character concentrate on the high-temperature properties of the superionic phase, and the reports on the low-temperature behavior are limited.	
	Recently, inelastic neutron scattering on Ag$_8$SnSe$_6$ has revealed that the phonon spectrum of the material is unusually broadened already at several 10 K, which is much lower than the temperature the superionic conduction occurs \cite{AgSnSe}.
	This fact implies that a characteristic lattice vibration with very high anharmonicity exists in the superionic conductors, and simultaneously, the low-temperature behavior includes essential information to understand the thermal properties of these materials.
	Broadening of the phonon signal far below the transition has also been observed in AgCrSe$_2$ \cite{kappa4}, although the details have not been mentioned.

\begin{figure}[t]
\begin{center}
\includegraphics[width=55mm]{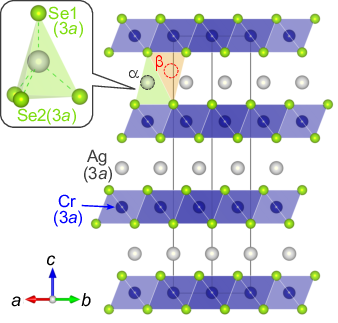}
\caption{\label{str} 
	Crystal structure of the low-temperature phase of AgCrSe$_2$ (a space group $R3m$). 
	The crystal structures throughout this paper are all visualized using VESTA \cite{VESTA}.
	There are two tetrahedral seats for Ag atoms, as indicated by $\alpha$ and $\beta$, one of which is preferentially occupied at temperatures below the $T_{\rm od}$. 
}
\end{center}
\end{figure}   

	Here, we focus on the layered chromium selenides $M$CrSe$_2$ ($M$ = Cu and Ag) and their possibility of introducing atomic disorder to the superionic species.
	The reported thermal conductivity at room temperature is $\sim$0.8 and 0.4 W K$^{-1}$ m$^{-1}$ for $M$ = Cu and Ag, respectively \cite{kappa1,kappa2,kappa3,kappa4,kappa5}.
	At high temperatures, they crystallize in a rhombohedral structure (a space group $R\bar{3}m$) with face-sharing CrSe$_6$ octahedral layers stacked along the $c$-axis direction of the trigonal cell.
	The succession of the CrSe$_6$ layers forms elongated tetrahedral sites with equivalent two seats, which are indicated as $\alpha$ and $\beta$ in Fig. \ref{str}.
	The $M$ ions expediently occupy the two seats at a ratio of 1:1 \cite{structure1} but are delocalized and mobile by hopping between them, resulting in the superionic conduction.
	The structure undergoes an order-disorder transition for the occupancy of $M$ ions at 365 and 475 K for CuCrSe$_2$ and AgCrSe$_2$, respectively, below which one of the seats is preferentially occupied  \cite{kappa1,structure2,Gagor}.
	Consequently, the inversion center is broken, and the crystal transforms into the low-temperature structure of a space group $R3m$.   
	This transition accompanies a sharp jump in the specific heat and a discontinuity in the lattice constant \cite{Gagor,INS-Ag-LA,INS_CuCrSe2,topical_review}, indicating that the transition is of the first-order character.

	Several attempts have been made to clarify the extremely low thermal conductivity of this system, and the proposed mechanisms include the localized vibrations by quasi-2D potential \cite{kappa4}, and the strong anharmonicity \cite{INS-Ag-PNAS}.
	According to phonon calculations and inelastic neutron scattering, CuCrSe$_2$ and AgCrSe$_2$ possess transverse acoustic (TA) branches with flat dispersions exhibiting a density-of-states (DOS) peak at 8 and 4 meV, respectively \cite{INS_CuCrSe2,INS-Ag-LA,INS-Ag-PNAS}. 
	Site-projected DOS calculations have revealed that the TA modes are characterized as the in-plane vibration of the conducting Cu and Ag ions.	
	It should be noted that the energy of the DOS peak of AgCrSe$_2$ is approximately half of that of CuCrSe$_2$.
	Here, the thermal conductivity of AgCrSe$_2$ is also approximately half of that observed in CuCrSe$_2$ at room temperature.
	This coincidence implies that the in-plane vibration of Ag and Cu ions is crucial for the low thermal conductivity in this system.
	The flat dispersions across the $k$-space should generate highly localized vibrations owing to the very low group velocities and have also been believed to contribute to the ultralow $\kappa_{\rm lat}$ \cite{INS-Ag-PNAS}.

	On the other hand, the longitudinal acoustic (LA) branch has been characterized as the vibration along the $c$ axis and exhibits the DOS peak at 12 and 10 meV for $M$ = Cu and Ag, respectively.
	These longitudinal vibrations have also been discussed as the candidate: inelastic neutron scattering (INS) experiments using AgCrSe$_2$ powder have reported that the LA mode survives even at high temperatures above the order-disorder transition temperature ($T_{\rm od}$), and in contrast, the TA modes are damped entirely by the diffusive Ag ions \cite{INS-Ag-LA}.
	Based on these results, the authors conclude that the robust LA mode causes the ``liquid-like'' low thermal conductivity.
	However, another INS experiment argues that the TA modes do coexist with the Ag diffusion in the superionic phase \cite{INS-Ag-PNAS}.

	From the structural viewpoint, the anisotropic displacement parameter $U_{11}$ for the $M$ site has been reported to be an order of magnitude larger than $U_{33}$ in AgCrSe$_2$ \cite{INS-Ag-LA}.
	The highly anisotropic displacement parameters have also been reported in AgCrS$_2$ \cite{Whittingham,Gerards,Damay1}.
	These anisotropic values are attributed to the amplitudes of the in-plane and out-of-plane vibrations of the $M$ atoms, respectively.
	In addition, the reported value of $U_{11}$ for AgCrSe$_2$ is approximately 0.1 \AA$^2$ at room temperature, which is larger than that for  CuCrSe$_2$ by a factor of two \cite{Gagor,INS-Ag-LA}.
	These facts raise questions if the magnitude of $U_{11}$ itself is important in reducing $\kappa$ or if other factors such as anharmonicity have a major contribution.
	In addition, it has been reported that the $U_{11}$ of the Ag ions maintains a large value even at 4 K \cite{structure1}, suggesting that the in-plane vibration of Ag ions may be preserved down to near the absolute zero temperature. 
	Based on these backgrounds, we aim to clarify the role of the Ag-site disorder on thermal conductivity by applying atomic substitution of 11$^{\rm th}$ group elements, Cu and Au, for the Ag site.

\section{Experimental}
   Powder samples of Ag$_{1-x}M_x$CrSe$_2$ ($M$ = Cu and Au) were synthesized using a conventional solid-state reaction. 
   Ag (Rare Metallic, 99.9\%, powder), Cu (Rare Metallic, 99.9\%, powder), Au (Nilaco, 99.99\%, powder), Cr (Rare Metallic, 99.9\%, powder), and Se (Kojundo Chemical Laboratory, 99.9\%, powder) were mixed at a molar ratio of 1-$x$ : $x$ : 1 : 2, uni-axially pressed into a pellet, and then put in an evacuated silica tube. 
   Before mixing, Cu powder was heated at 200$^{\circ}$C for one hour in flowing H$_2$ gas to reduce copper oxides.
   The tube was first heated to 200$^{\circ}$C in one hour and kept for 6 hours. 
   The tube temperature was raised to 900$^{\circ}$C at a ratio of 350$^{\circ}$C/h, kept for 24 hours, and then cooled in a furnace at a ratio of approximately 150$^{\circ}$C/h. 
   Sample purity was confirmed using a laboratory x-ray diffraction (Cu-K$\alpha$) 
   prior to the physical property measurements.
   Differential scanning calorimetry (DSC) was performed to determine the $T_{\rm od}$.
   Samples for physical property measurements were prepared as follows. 
   The powder sample was uni-axially pressed into a pellet, pressed again in hydrostatic pressure, put in an evacuated silica tube, and then sintered at 900$^{\circ}$C for 20 hours.
   Samples were cut into a rectangular shape for thermal conductivity, electrical resistivity, and specific heat measurements.

   Electrical resistivity and specific heat were measured using a four-probe method and the relaxation method, respectively, in a physical property measurement system (PPMS, Quantum Design).
   Thermal conductivity was measured using the steady-state method in the PPMS, with a commercial sample holder. 
   The values of the cold and hot thermometers were read by the built-in instruments in the PPMS, and the heater power was controlled using external devices connected to the heater terminals.
   Synchrotron x-ray diffraction was performed at the BL02B2 and BL19B2 beamlines at SPring-8.

\section{Results and discussion}

	Figs. S1(a) and S1(b) \cite{SM} display the powder x-ray diffraction patterns for $M$ = Cu and Au ($x\leq0.05$), respectively, obtained using the laboratory x-ray diffractometer.
	While the samples remain in a single phase up to $x=0.03$, a secondary phase of Ag$_2$Se or Se appears at compositions of $x\geq0.04$ and $x=0.05$ for $M$ = Cu and Au, respectively.
	These secondary phases can occur independently of Cu or Au addition and are not due to an excess amount of substitution.
	However, their occurrence is minimized as much as possible.

\begin{figure}[t]
\begin{center}
\includegraphics[width=53mm]{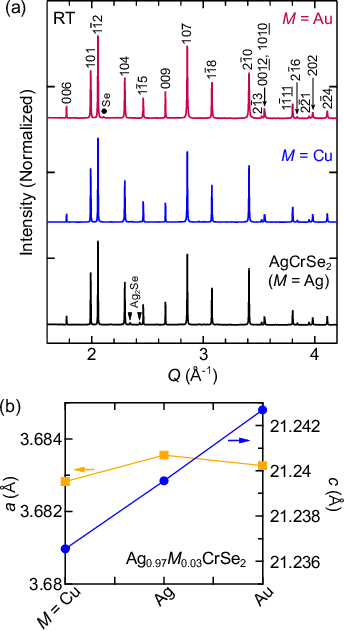}
\caption{\label{sXRD} 
	(a) Synchrotron x-ray diffraction patterns of AgCrSe$_2$ and Ag$_{0.97}M_{0.03}$CrSe$_2$ ($M$ = Cu, Au) powders obtained at room temperature.
	Data were collected at the BL19B2 beamline.
	The incident x-ray energy is set at 24 keV, and the calibrated wavelength $\lambda$ is 0.51689 \AA. Filled triangles indicate Ag$_2$Se, and a filled circle represents Se as secondary phases.
	(b) Lattice constants $a$ and $c$ obtained from the diffraction patterns. The error bars are within the range of the marker size.
}
\end{center}
\end{figure}

	Fig. \ref{sXRD}(a) shows the synchrotron x-ray diffraction (sXRD) patterns of AgCrSe$_2$ and Ag$_{0.97}M_{0.03}$CrSe$_2$ ($M$ = Cu and Au) measured at room temperature.
	Samples used for the synchrotron experiments were different from those shown in Fig. S1 and include a small amount of the secondary phase.
	In addition, they were prepared by carefully gathering a very small amount of originally fine crystal grains, which were obtained without grinding the sample after reaction, and placing them in a capillary with a diameter of 0.2 mm. 
	The lattice constants obtained from the sXRD are summarized in Fig. \ref{sXRD}(b). 
	The lattice constant $a$ is insensitive both for Cu and Au addition and shows an almost constant value. 
	The lattice constant $c$ varies systematically depending on the ionic size of the $M$ site atom, indicating the successful substitution of these elements for the Ag site.

\begin{figure}[t]
\begin{center}
\includegraphics[width=85mm]{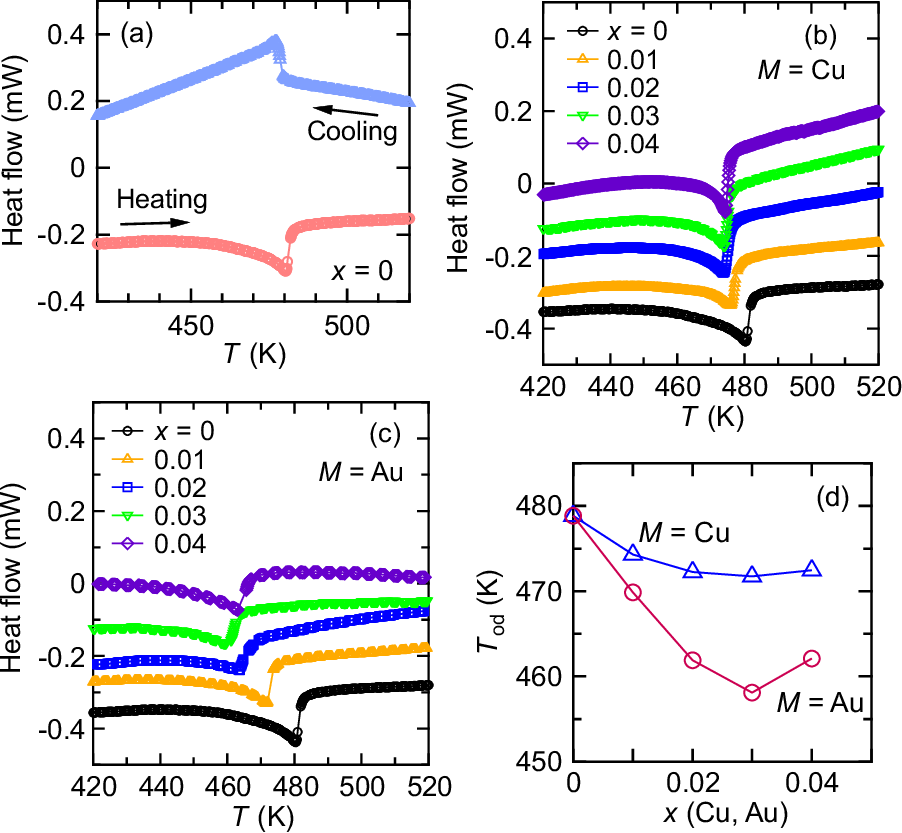}
\caption{\label{DSC} 
(a) DSC curves of $x=0$ obtained during heating and cooling procedures. Because of the first-order character of the order-disorder structural phase transition, there is a small difference between the peak positions of the two curves. 
(b) and (c) show the DSC curves of $M$ = Cu and Au powder samples, respectively, in a chemical formula Ag$_{1-x}M_{x}$CrSe$_2$. The curves were obtained during a heating procedure at a rate of 10 K / min. 
Data are offset vertically for clarity.
(d) $T_{\rm od}$ of Ag$_{1-x}M_{x}$CrSe$_2$ ($M$ = Cu and Au) powder samples. 
The $T_{\rm od}$ was determined as a midpoint of the peak temperatures of the heating and cooling curves.
}
\end{center}
\end{figure}

	Fig. \ref{DSC}(a) shows the DSC curves of $x=0$ sample obtained during heating and cooling procedures.
	A peak is observed in each curve approximately at 480 K, indicating the order-disorder transition of Ag ions.
	Because the peak positions slightly differ from each other due to the first-order nature of the structural phase transition, 
	the $T_{\rm od}$ is defined as the midpoint between the peak positions observed in the two curves.
	$T_{\rm od}$ of $x=0$ determined in this way is 479 K, which agrees well with the reported value \cite{structure2}.

	Figs. \ref{DSC}(b) and \ref{DSC}(c) represent the DSC curves for $M$ = Cu and Au, respectively, obtained during a heating procedure. 
	All the DSC data, including cooling procedures, are presented in Fig. S4 \cite{SM}.
	 Fig. \ref{DSC}(d) shows the determined $T_{\rm od}$ plotted as a function of $x$. 
	The $T_{\rm od}$ systematically decreases with increasing the substitution levels of Cu and Au up to $x$ = 0.03. 
   Further decrease in the $T_{\rm od}$ is not observed at $x$ = 0.04 both for Cu and Au, indicating that the solid solubility limit is approximately at $x$ = 0.03 for both elements.
   	The decrease in the $T_{\rm od}$ by Au substitution is also confirmed using synchrotron x-ray diffraction with varying temperatures, as shown in Figs. S2 and S3 \cite{SM}.   
   Figs. S5(a) and S5(b) \cite{SM} show the DSC curves of heating and cooling procedures, respectively, obtained using another series of the Cu-substituted samples from those plotted in Fig. \ref{DSC}(b).
   The $T_{\rm od}$ determined from these curves are compared in Fig. S5(c) with the samples described above.
   The $x$ variation of $T_{\rm od}$ shows a good reproducibility, as shown in Fig. S5(c), and the uncertainty in the $T_{\rm od}$ is found to be approximately 1 K.

	The decrease in the transition temperature indicates that the structure of the low-temperature phase becomes unstable, making the structure easier to transition to the high-temperature phase where the Ag ions are in the disordered state. 
	Additionally, the transition temperature decreases with both Cu substitution and Au substitution. 
	This fact suggests that the structural phase transition in this system is dominated not by the size of the ions occupying the Ag site, but by the degree of disorder at the Ag site.
	In the Au-substituted samples, the decrease in $T_{\rm od}$ is more than 20 K, whereas the decrease in $T_{\rm od}$ in the Cu-substituted samples is smaller, around 10 K. 
	It is believed that the Au substitution makes the structure more unstable and easier to transform to the high-temperature phase than the Cu substitution because Au tends to form bonds less easily with surrounding Se atoms compared to Cu.

\begin{figure}[t]
\begin{center}
\includegraphics[width=60mm]{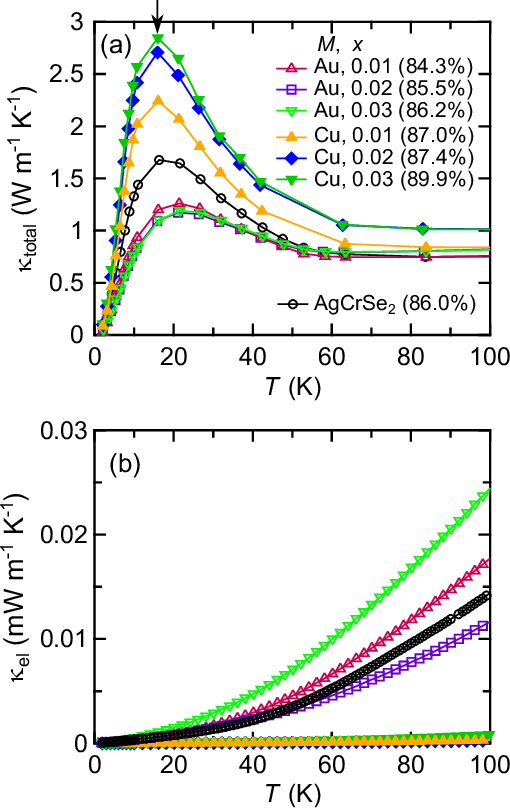}
\caption{\label{kappa} 
(a) Total thermal conductivity ($\kappa_{\rm total}$) of Ag$_{1-x}M_{x}$CrSe$_2$ polycrystalline samples ($M$ = Cu and Au; $x=0$--0.03). The numbers in the parentheses indicate the relative density of the samples.
(b) Electronic contribution to the thermal conductivity ($\kappa_{\rm el}$) obtained using electrical resistivity shown in Fig. S8(a) \cite{SM} and Lorenz number. 
}
\end{center}
\end{figure}

	Temperature dependence of total thermal conductivity ($\kappa_{\rm total}$) is displayed in Fig. \ref{kappa}(a). 
	The numbers in parentheses indicate the relative density of the samples used for the measurement.
	Open black circles represent the $\kappa_{\rm total}$ of pristine AgCrSe$_2$.
	It has been known that the relative density affects the value of thermal conductivity.
	Fig. S6(a) \cite{SM} compares the $\kappa_{\rm total}$ measured for samples with different relative densities.
	As shown in Fig. S6(a), a small difference in the relative density does not affect the thermal conductivity at low temperatures below $\sim$50 K, although it affects the values at high temperatures.
	Additionally, as shown in Fig. S6(b), the reproducibility of the data has been confirmed using samples with the same composition and density.
	These results mean that, at temperatures below 50 K, it is possible to discuss the trends among these samples.
	The magnitude of $\kappa_{\rm total}$ for $x=0$ shown in Fig. \ref{kappa}(a) agrees well with the reported values \cite{kappa4} measured for the polycrystalline samples that were prepared using a sintering method similar to this study.

	In Fig. \ref{kappa}(a), as the Cu-substitution level $x$ increases, the $\kappa_{\rm total}$ systematically increases.
	In contrast, the $\kappa_{\rm total}$ systematically decreases as the Au-substitution level $x$ increases.
	No systematic change is observed in the high-temperature region above 100 K, as shown in Fig. S7  \cite{SM}, which is probably due to the difference in the relative density between the samples.
	The electronic contribution to the thermal conductivity ($\kappa_{\rm el}$) is evaluated by using electrical resistivity shown in Fig. S8(a) \cite{SM} and the Lorenz number, $L=2.45 \times 10^{-8}$ [W $\Omega$ K$^{-2}$]. 
	The obtained values of $\kappa_{\rm el}$ are shown in  Fig. \ref{kappa}(b).
	They are low enough in all the samples that the $\kappa_{\rm total}$ shown in Fig. \ref{kappa}(a) can be assumed to have a negligibly small electronic contribution.
	The lattice contribution ($\kappa_{\rm lat}$) obtained by subtracting $\kappa_{\rm el}$ from $\kappa_{\rm total}$ is displayed in Fig. S8(b).

	The in-plane electrical resistivity ($\rho_{ab}$) of AgCrSe$_2$ single crystals has been reported to be 9 m$\Omega$ cm at 300 K \cite{rho_AgCrSe2}.
	Although there is no report in the out-of-plane electrical resistivity ($\rho_{c}$) for AgCrSe$_2$, the electrical resistivity of this system has been known to be highly anisotropic; the value of $\rho_{c}$ is three orders of magnitude greater than that of $\rho_{ab}$ in the sulfur system, AgCrS$_2$ \cite{rho_AgCrSe2}. 
	In polycrystalline samples, the values of electrical resistivity generally include both the in-plane and out-of-plane components.
	Indeed, the $x=0$ sample shown in Fig. S8(a) exhibits the expected values as the order of magnitude intermediate between the in-plane and out-of-plane values.

	As shown in Fig. S8(a), the values of electrical resistivity of the Au-substituted sample is almost the same as that of $x=0$.
	On the other hand, the values of electrical resistivity of the Cu-substituted sample is one order of magnitude greater than those obtained for $x=0$.	
	The reason for this significant increase in the electrical resistivity by the Cu substitution can be explained as follows. 
	First, since the Seebeck coefficient is positive across all temperature ranges \cite{kappa4}, the electrical conduction carriers are holes. 
	According to the electronic state calculations \cite{TN_Baenitz}, AgCrSe$_2$ has bands composed of Cr and Se near the Fermi level.
	However, the density of states at the Fermi level is very small, suggesting the presence of small hole pockets.
	Given that the amount of the Cu substitution is only a few percent, the rigid band model can be applied, assuming that the band shape does not change with the Au or Cu substitution. 
	 Additionally, Ag and Au can be considered to be monovalent in AgCrSe$_2$.
	 However, Cu generally prefers a divalent state rather than monovalent.
	 Therefore, Cu substitution acts as electron doping, and even a small amount of substitution significantly reduces the number of hole carriers, resulting in the large increase in the electrical resistivity.

	As marked by an arrow in Fig. \ref{kappa}(a), all the samples exhibit a peak at $\sim$15 K, which is one of the characteristics generally observed in crystalline samples: in the kinetic formulation, lattice thermal conductivity ($\kappa_{\rm lat}$) is given by the product of lattice specific heat ($C_{\rm lat}$), average velocity of phonon ($\nu$), and phonon mean free path ($\ell$) as $\kappa_{\rm lat} = 1/3 C_{\rm lat} \nu \ell = 1/3 C_{\rm lat} \nu^2 \tau$.
	Here, $\tau$ is the phonon lifetime.
	At the lowest temperatures, $\ell$ has a constant value, and thus the $\kappa_{\rm lat}$ behaves as $T^3$ dominated by the temperature dependence of $C_{\rm lat}$. 
   As temperature increases, the number of phonons increases, which leads to $\kappa_{\rm lat} \propto \exp(\Theta / T)$, where $\Theta$ is a temperature on the order of Debye temperature \cite{AM_book}. 
   As a result, $\kappa_{\rm lat}$ generally exhibits a peak at a temperature where the dominant temperature dependence changes. 
   Because the atomic mass of Au is larger than that of Ag, the reduced $\kappa$ in the Au-substituted samples might be intuitive; the heavy Au ions scatter phonons more effectively than the Ag ions in the pristine sample.

\begin{figure}[t]
\begin{center}
\includegraphics[width=85mm]{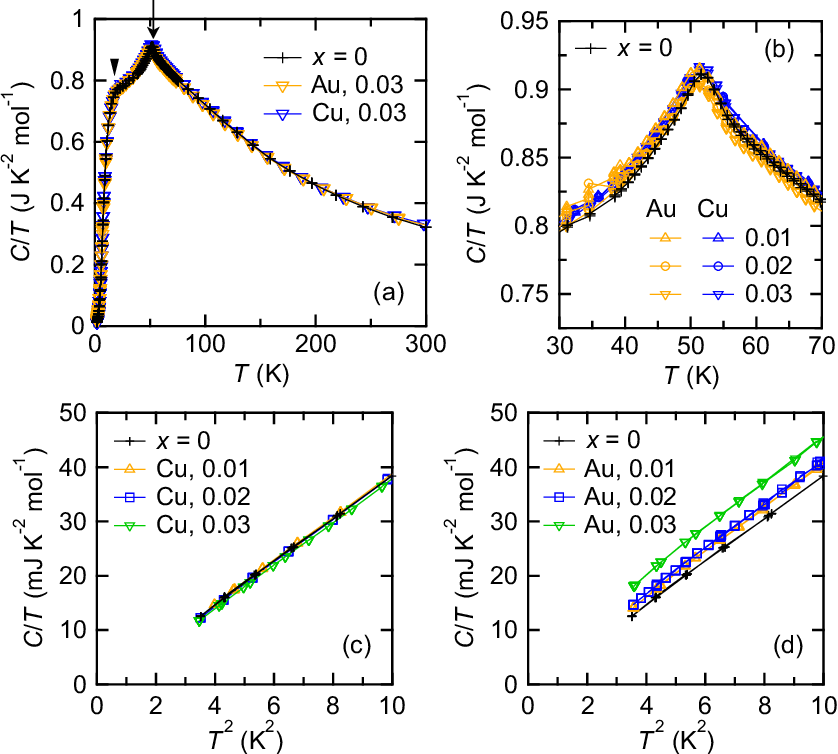}
\caption{\label{Cp} 
(a) Specific heat of the samples. The anomaly at $T_{\rm N}$ = 52 K indicated by an arrow is attributed to a magnetic transition \cite{structure1}.
A large hump is also observed approximately at 20 K, as shown by a triangle, probably caused by the residual entropy due to disorder components of spin fluctuations. 
(b) Magnified view of the temperature region around the magnetic transition temperature. 
(c) and (d) display the $C/T$ vs. $T^2$ plots of the specific heat of $M$ = Cu and Au, respectively. 
}
\end{center}
\end{figure}

 \begin{table*}[t] 
 \caption{\label{Uij} Structural parameters and anisotropic atomic displacement parameters $U_{ij}$ (\AA$^2$) for AgCrSe$_2$ and Ag$_{0.97}M_{0.03}$CrSe$_2$ ($M$ = Cu and Au). The nominal compositions were used for the refinement. 
}
 \begin{center}
 \begin{tabular}{cccccccccccc} \hline\hline 
 \multicolumn{11}{c}{AgCrSe$_2$, $a$ = 3.68355(1) \AA, $c$ = 21.23955(6) \AA, $R_{\rm wp}$ = 5.53\%}\\ \hline
 Atom & site & $x$ & $y$ & $z$ & $U_{\rm iso}$ & $U_{11}$ & $U_{22}$ & $U_{33}$ & $U_{12}$ & $U_{13}$ & $U_{23}$  \\ \hline 
 \vspace{1mm}
 Cr & $3a$ & 0 & 0 & 0.000(9) & 0.0075(4) & & & & & \\ 
 Se1 & $3a$ & 0 & 0 & 0.269(9) & 0.0042(4) & & & & & \\ 
 Se2 & $3a$ & 0 & 0 & 0.733(9) & 0.0086(5) & & & & & \\ 
 Ag & $3a$ & 0 & 0 & 0.153(8) & 0.0604(12) & 0.0841(13) & $=U_{11}$ & 0.0129(9) & $=1/2 \; U_{11}$ & 0 & 0 \\ \hline\hline
 \multicolumn{11}{c}{Ag$_{0.97}$Cu$_{0.03}$CrSe$_2$, $a$ = 3.68283(1) \AA, $c$ = 21.23655(6) \AA, $R_{\rm wp}$ = 5.40\%}\\ \hline
 Atom & site & $x$ & $y$ & $z$ & $U_{\rm iso}$ & $U_{11}$ & $U_{22}$ & $U_{33}$ & $U_{12}$ & $U_{13}$ & $U_{23}$  \\ \hline 
 \vspace{1mm}
 Cr & $3a$ & 0 & 0 & 0.000(8) & 0.0076(3) & & & & & \\ 
 Se1 & $3a$ & 0 & 0 & 0.269(8) & 0.0067(4) & & & & & \\ 
 Se2 & $3a$ & 0 & 0 & 0.733(8) & 0.0084(4) & & & & & \\ 
 Ag/Cu & $3a$ & 0 & 0 & 0.152(8) & 0.0568(10) & 0.0738(11) & $=U_{11}$ & 0.0228(9) & $=1/2 \; U_{11}$ & 0 & 0 \\  \hline\hline 
 \multicolumn{11}{c}{Ag$_{0.97}$Au$_{0.03}$CrSe$_2$, $a$ = 3.68327(2) \AA, $c$ = 21.24268(9) \AA, $R_{\rm wp}$ = 4.96\%}\\ \hline
 Atom & site & $x$ & $y$ & $z$ & $U_{\rm iso}$ & $U_{11}$ & $U_{22}$ & $U_{33}$ & $U_{12}$ & $U_{13}$ & $U_{23}$  \\ \hline 
 \vspace{1mm}
 Cr & $3a$ & 0 & 0 & 0.000(9) & 0.0067(4) & & & & & \\ 
 Se1 & $3a$ & 0 & 0 & 0.269(9) & 0.0058(5) & & & & & \\ 
 Se2 & $3a$ & 0 & 0 & 0.733(9) & 0.0071(5) & & & & & \\ 
 Ag/Au & $3a$ & 0 & 0 & 0.152(9) & 0.0618(12) & 0.0908(14) & $=U_{11}$ & 0.0037(8) & $=1/2 \; U_{11}$ & 0 & 0 \\  \hline\hline 
 \end{tabular}
 \end{center}
 \end{table*}

	For a closer examination of this, we conducted specific heat measurements.
	Fig. \ref{Cp}(a) displays the specific heat ($C$) of AgCrSe$_2$ and Ag$_{0.97}M_{0.03}$CrSe$_2$ ($M$ = Cu, Au) polycrystalline samples. 
	The values on the longitudinal axis are the specific heat divided by temperature.
	Each $C/T$ vs. $T$ curve exhibits a $\lambda$-type peak at 52 K, as indicated by an arrow in Fig. \ref{Cp}(a). This anomaly is attributed to a magnetic ordering of  Cr$^{3+}$ spins \cite{TN_Bongers,TN_Gautam}.
	As shown in Fig. \ref{Cp}(b), the $T_{\rm N}$ remains unaffected with the Cu or Au substitution because the changes in the lattice constants are very small.

	Notable feature is a hump observed around 20 K indicated by a triangle in Fig. \ref{Cp}(a). 
	According to M. Baenitz {\it et al.} \cite{TN_Baenitz}, there are several in-plane exchange interactions between Cr$^{3+}$ ions. 
	The nearest-neighbor interaction within the $ab$-plane is ferromagnetic, whereas the third-nearest-neighbor interaction is antiferromagnetic.
	The interaction between the planes is antiferromagnetic and significantly weaker compared to these interactions. 
	Because of the magnetic frustration effect due to the competing exchange interactions between the two major interactions, Cr$^{3+}$ has been known to form a cycloidal magnetic structure within the plane \cite{TN_Baenitz}.
	The observed hump might be from this cycloidal magnetic structure. 
	The neutron diffraction \cite{TN_Baenitz} revealed that the magnitude of the ordered magnetic moment of Cr$^{3+}$ at the lowest temperature is much smaller than the theoretical value predicted for $S=3/2$.
	This fact suggests the presence of structural disorder or the existence of spin fluctuations persisting to very low temperatures.
	Therefore, the hump observed around 20 K might reflect the residual entropy of spins from the disorder components or spin fluctuations. 
	Particularly, such humps are often observed in frustrated systems \cite{CuCrO2,PdCrO2}.

	Figs. \ref{Cp}(c) and \ref{Cp}(d) display the $C/T$ of Cu- and Au-substituted samples, respectively, as a function of $T^2$.
	The values of $C/T$ is almost unchanged by the Cu substitution, as shown in Fig. \ref{Cp}(c).
	On the contrary, the Au substitution increases the $C/T$.
	These results indicate that low-energy phonons, {\it i.e.}, the phonon DOS at low energies, increase in the Au-substituted AgCrSe$_2$.
	Because the increased number of phonons should scatter other phonons more frequently and reduce $\tau$, the increase in $C/T$ at low temperatures explains the reduced thermal conductivity of the Au-substituted AgCrSe$_2$.	
	The $T_{\rm N}$ and the temperature at which the hump is observed remain unchanged by the Cu and Au substitutions. Therefore, the increase in $C/T$ is attributed to the lattice contribution rather than the magnetic contribution.

\begin{figure*}[t]
\begin{center}
\includegraphics[width=170mm]{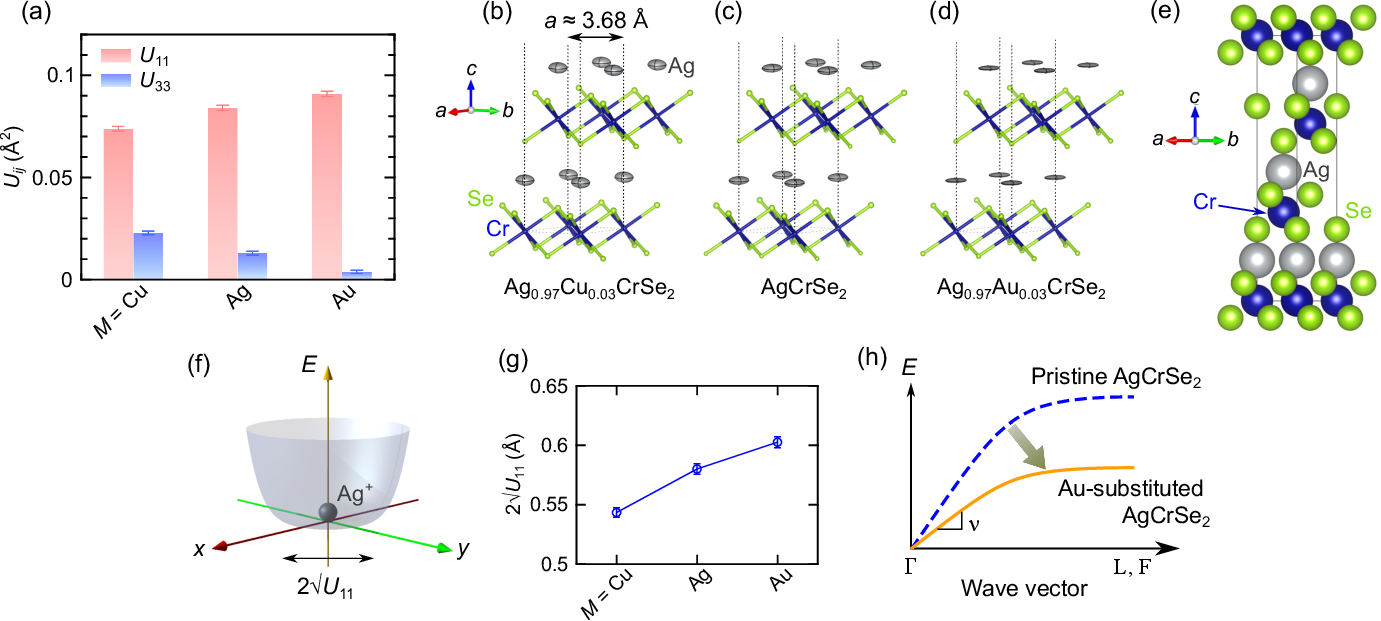}
\caption{\label{Uij_plot} 
(a) Anisotropic atomic displacement parameters $U_{11}$ and $U_{33}$ for AgCrSe$_2$ and Ag$_{0.97}M_{0.03}$CrSe$_2$ ($M$ = Cu and Au) samples. $M$ = Ag represents AgCrSe$_2$.
(b), (c) and (d) display the crystal structures at room temperature of Ag$_{0.97}$Cu$_{0.03}$CrSe$_2$, AgCrSe$_2$, and Ag$_{0.97}$Au$_{0.03}$CrSe$_2$, respectively, with the anisotropic displacement ellipsoids drawn at 80\% probability level for the Ag/Cu, Au atoms.
(e) Crystal structure of AgCrSe$_2$ where the space filling is visualized using ionic radii. 
(f) Schematic image for a flat-bottomed potential well generated around each Ag ion. The value of $2\sqrt{U_{11}}$ is a measure of the width of the well indicated by a bothsided arrow.
(g) The values of $2\sqrt{U_{11}}$ plotted for $M$ = Cu, Ag, and Au.
(h) Schematic diagram of the TA phonon dispersions. The curve of the pristine AgCrSe$_2$ refers to the results of the phonon calculation \cite{INS-Ag-PNAS}. $L$ and $F$ represent the symmetry points at the first Brillouin zone boundary of the rhombohedral cell, and the wave vectors lie approximately within the Ag ion-conducting plane.
$\nu$ in the formulation of $\kappa_{\rm lat}=1/3 C_{\rm lat} \nu \ell$ corresponds to the slope of the long-wavelength linear dispersion of the acoustic branch.
}
\end{center}
\end{figure*}

	To delve deeper into the reduced thermal conductivity in the Au-substituted samples, the Rietveld analysis was performed for the pristine, and Ag$_{0.97}M_{0.03}$CrSe$_2$ ($M$ = Cu and Au) samples.
	Although a small amount of secondary phases has been observed in the profiles shown in Fig. \ref{sXRD}(a), their quantity is minimal and does not affect the analysis.	
	The nominal compositions are used for the refinement.
	All the fittings yielded low $R_{\rm wp}$ factors and reproduce well the experimentally obtained diffraction patterns, as shown in Figs. S9(a)--(c) \cite{SM}.

	Table \ref{Uij} summarizes the refined structural parameters for the three samples.
	Anisotropic atomic displacement parameters, $U_{ij}$, are refined only for the Ag site because the refined values of the isotropic displacement parameters, $U_{\rm iso}$, were found to be very small at the Cr and Se sites.
	The atomic positions show no significant differences among the three samples.
	On the other hand, the parameters $U_{11}$ and $U_{33}$ of the Ag site exhibit different values depending on the samples, which are plotted in Fig. \ref{Uij_plot} (a) in the order of $M$ = Cu, Ag, and Au. 
	Here, $M$ = Ag represents the pristine AgCrSe$_2$.

	As displayed in Fig. \ref{Uij_plot}(a), $U_{11}$ systematically increases, while $U_{33}$ decreases, following the sequence of $M$ = Cu, Ag, and Au.
	The variation of $U_{11}$ and $U_{33}$ is visualized in Figs. \ref{Uij_plot}(b)--(d) using anisotropic displacement ellipsoids drawn at 80\% probability level.
	Generally, atomic displacement parameters reflect the degree of static or dynamic disorder of the atoms. 
	In this system, the Ag atoms exhibit the characteristic TA and LA modes with low energies vibrating within the $ab$-plane and along the $c$ axis, respectively \cite{INS-Ag-PNAS}. 
	Therefore, the values of the obtained $U_{11}$ and $U_{33}$ mainly reflect the amplitude of the local atomic vibrations. 
	As these localized vibration modes can effectively scatter other phonons, they are anticipated to reduce thermal conductivity.
	Additionally, larger vibration amplitudes would have greater scattering effects. 
	Returning back to Fig. \ref{kappa}(a), the thermal conductivity systematically decreases with increasing Au substitution.
	Furthermore, $U_{11}$ increases with Au substitution.
	Therefore, the increase in $U_{11}$ is responsible for the reduced thermal conductivity in the Au-substituted samples.

	The composition dependent $U_{11}$ and $U_{33}$ 
	indicate that the shape of the vibrational potential at the Ag site is altered by the Cu and Au substitutions. 
	The ionic radii are in the order of Cu $<$ Ag $<$ Au, and the Ag site is surrounded by Se.
	The filling space for Ag ions along the $c$-axis is small, as shown in Fig. \ref{Uij_plot}(e). 
	On the other hand, there is larger space in the $ab$-plane, making the ions inherently easier to vibrate within the plane. 
	Therefore, the change in the $U_{11}$ and $U_{33}$ due to the Cu and Au substitution can be explained as follows: 
	The size of the ionic radius relative to the Ag ion significantly affects the ease of vibration along the $c$-axis direction, while the vibration amplitude is large within the $ab$-plane.
	As a result, the Au substitution suppresses the interplanar vibrations and enhances the in-plane vibrational amplitude, thereby modifying the potential energy landscape associated with the vibrations.
	Assuming a uniform distribution of the 3\% substituent elements across the Ag plane, the average distance between them within the $ab$-plane is approximately 19 \AA.
	There are only three unit cells apart along the [110] direction, and it is sufficient for the local modulation of the vibrational potential to significantly affect the overall vibrational state.

	According to Raman scattering experiments \cite{Biswas}, the localized low-energy vibrations in AgCrSe$_2$ are concluded as anharmonic vibrations which interact with higher optical modes to achieve low thermal conductivity. 
	Fig. \ref{Uij_plot}(f) depicts the schematic image for the anharmonic potential well at the Ag sites.
	Here, the square root of the $U_{11}$ corresponds to the amplitude of the in-plane vibrations at the Ag sites. 
	Therefore, the value of $\sqrt{U_{11}}$ multiplied by a factor of 2 is a measure of the width of the potential well indicated by the double-sided arrows in Fig. \ref{Uij_plot}(f).
	Fig. \ref{Uij_plot}(g) plots the values of $2\sqrt{U_{11}}$ for $M$ = Cu, Ag, and Au.
	It increases in the order of $M$ = Cu, Ag, and Au, reaching up to 16.4\% of the Ag-Ag distance ($=$ lattice constant $a$) for $M$ = Au. 
	These results indicate that the anharmonicity is enhanced by the Au substitution and contributes to the reduction of the lattice thermal conductivity
	 through increased phonon-phonon scattering.
	The contribution of anharmonic vibrations to the reduction of thermal conductivity has also been observed in several compounds \cite{rattling1,tetrahedrite}.

	The enhancement in the in-plane vibrational amplitude should also decrease the force constants between the adjacent Ag/Au ions, {\it i.e.}, decrease in the Ag-Ag bonding strength, which should cause softening of the TA branch, as shown in Fig. \ref{Uij_plot}(h).
	The reduced slope of the TA mode indicates a lower phonon velocity, $\nu$, which directly contributes to the reduction of lattice thermal conductivity, as $\kappa_{\text{lat}} \propto \nu^2 \tau$.
	Simultaneously, the number of low-energy phonons would increase, which explains the trend of $C/T$ shown in Fig. \ref{kappa}(c), and further contributes to the suppression of thermal transport.

Ultra-low thermal conductivity in ionic conductors has long remained a mystery in the field of thermoelectrics, and the relationship between atomic dynamics and thermal transport has not been clearly understood, as discussed in the Introduction. 
The main achievement of this study is not merely the observation of reduced thermal conductivity due to Au substitution, but rather identification of enhanced anharmonicity in the ion-conducting species as the underlying origin of the ultra-low thermal conductivity. 
This anharmonicity leads to an increased number of low-energy phonons, which further contributes to the suppression of thermal transport.
These insights are broadly applicable to the thermal behavior not only of other superionic conductors, but also of compounds with fluctuating sublattices \cite{InTe, TlInTe2}, extending beyond the AgCrSe$_2$ system itself. In this context, the progress described in this study offers a solid foundation for advancing the use of ionic conductors in thermoelectric applications. We believe our results provide meaningful guidance for future strategies in lattice-related functional materials.

\section{Conclusions}
	Thermal conductivity, electrical resistivity, specific heat, and structural parameters have been characterized for Ag$_{1-x}M_x$CrSe$_2$ ($M$ = Cu and Au).
	The Au substitution for the Ag site systematically decreases the thermal conductivity with increasing $x$.
	On the contrary, the Cu substitution enhances the thermal conductivity.
	Specific heat measurements reveal that Au substitution leads to an increase in the low-energy phonons.
	Powder x-ray structure refinement further demonstrates that Au substitution increases the in-plane atomic displacement parameter $U_{11}$.
	This clearly indicates an enhancement in the amplitude of in-plane vibrations of Ag atoms, which reflects increased anharmonicity in their vibrational potential.
	At the same time, the large vibrational amplitude also indicates a reduction in the force constants between Ag atoms, which results in a decrease in the phonon velocity of the corresponding acoustic mode.
	The increase in the low-energy phonons is likely caused by the reduction in these force constants.
	Both the increase in the low-energy phonons and the enhancement of in-plane vibrational amplitude contribute to a reduction in the phonon lifetime, and thus govern the ultra-low thermal conductivity of this system.
	This anharmonicity-based scenario is generally applicable to other superionic conductors, and we believe that our findings represent a solid step forward for future research on thermoelectric materials.

\begin{acknowledgements}
The authors thank Dr. K. Osaka for his support to conduct the experiments at BL19B2.
The synchrotron radiation experiments were performed at BL02B2 (Proposal No. 2022A2077) and BL19B2 of SPring-8. 
This work was partially supported by an SDGs Research Project of Shimane University. 
J.M. and J.F.Z. thank the financial support from the National Science Foundation of China (No. 12334008).
\end{acknowledgements}


\end{document}



\title{{\small \rm Supplemental Material} \\
\vspace{3mm}
Impact of in-plane disorders on the thermal conductivity of AgCrSe$_2$}


\author{S.~Izumi}
\affiliation{Department of Materials Science, Osaka Metropolitan University, Osaka 599-8531, Japan}

\author{Y.~Ishii}
\email{yishii@mat.shimane-u.ac.jp}
\affiliation{Co-Creation Institute for Advanced Materials, Shimane University, Shimane 690-8504, Japan}

\author{J.~Zhu}
\affiliation{School of Physics and Astronomy, Shanghai Jiao Tong University, Shanghai 200240, China}

\author{T.~Sakamoto}
\affiliation{Department of Materials Science, Osaka Metropolitan University, Osaka 599-8531, Japan}

\author{S.~Kobayashi}
\affiliation{Japan Synchrotron Radiation Research Institute (JASRI),  Hyogo 679-5198, Japan}

\author{S.~Kawaguchi}
\affiliation{Japan Synchrotron Radiation Research Institute (JASRI), Hyogo 679-5198, Japan}

\author{J.~Ma}
\affiliation{School of Physics and Astronomy, Shanghai Jiao Tong University, Shanghai 200240, China}

\author{S.~Mori}
\affiliation{Department of Materials Science, Osaka Metropolitan University, Osaka 599-8531, Japan}




\maketitle


\section{Powder x-ray diffraction experiment}

\subsection{XRD patterns obtained using Cu-K$\alpha$ radiation}

\begin{figure}[h]
\begin{center}
\includegraphics[width=120mm]{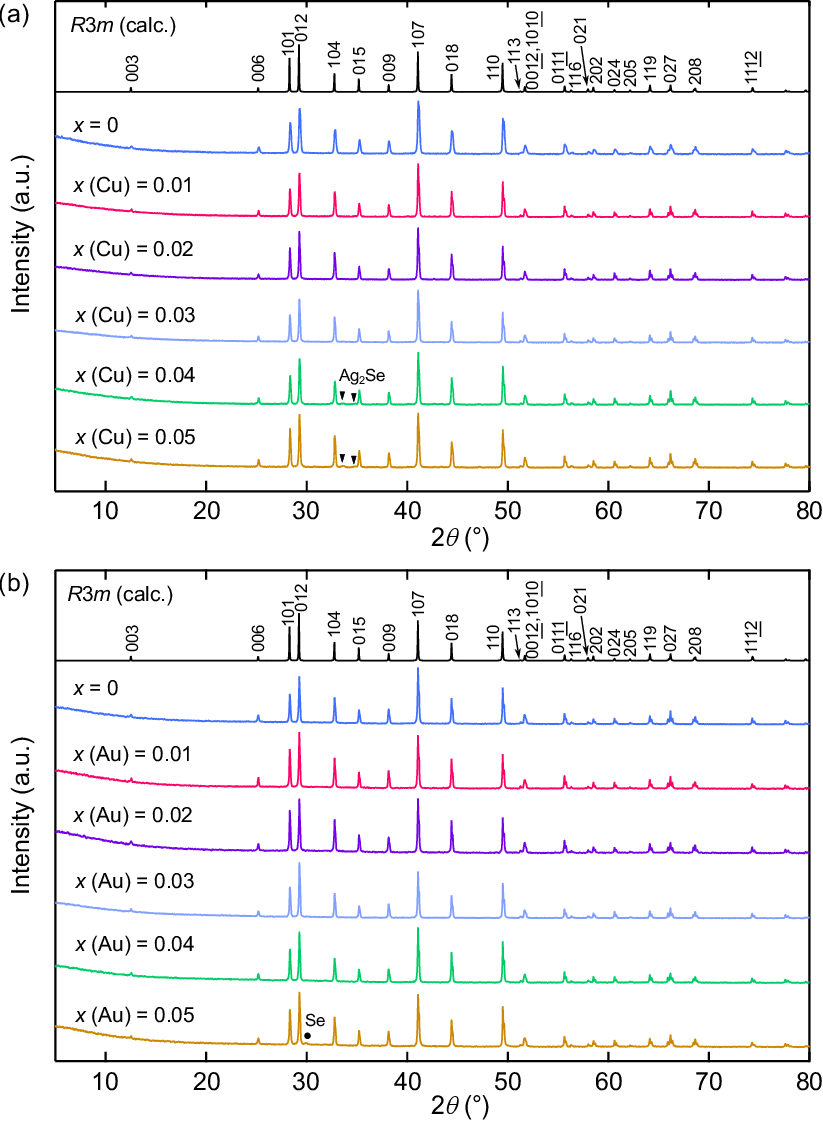}
\caption{\label{pxrd} 
Powder x-ray diffraction patterns of (a) $M$ = Cu and (b) $M$ = Au 
with $x\leq0.05$ obtained using Cu-K$\alpha$ radiation.}
\end{center}
\end{figure}

	Figs. \ref{pxrd}(a) and \ref{pxrd}(b) display the powder x-ray diffraction patterns for $M$ = Cu and Au ($x\leq0.05$), respectively, obtained using the laboratory x-ray diffractometer.
	 The calculated x-ray diffraction profile for the room temperature structure is also shown.
	 While the samples consist of a single phase up to $x=0.03$, secondary phases are observed at compositions of $x\geq0.04$ and $x=0.05$ for $M$ = Cu and Au, respectively.
	These are the same as those observed in the synchrotron x-ray diffraction patterns of $x$ = 0 and $x$(Au) = 0.03, as shown in Fig. 2(a).

	Due to the extremely long layered structure of this material along the $c$-axis, preparing samples for powder XRD is very challenging. As a result, it was difficult to obtain reliable lattice constants from the laboratory XRD diffraction experiments.
	For the synchrotron x-ray diffraction, the samples were prepared by carefully gathering a very small amount of originally fine crystal grains, which were obtained without grinding the sample after reaction, and placing them in a capillary with a diameter of 0.2 mm. 

\subsection{Order-disorder transition confirmed by temperature-dependent x-ray diffraction}

\begin{figure}[h]
\begin{center}
\includegraphics[width=160mm]{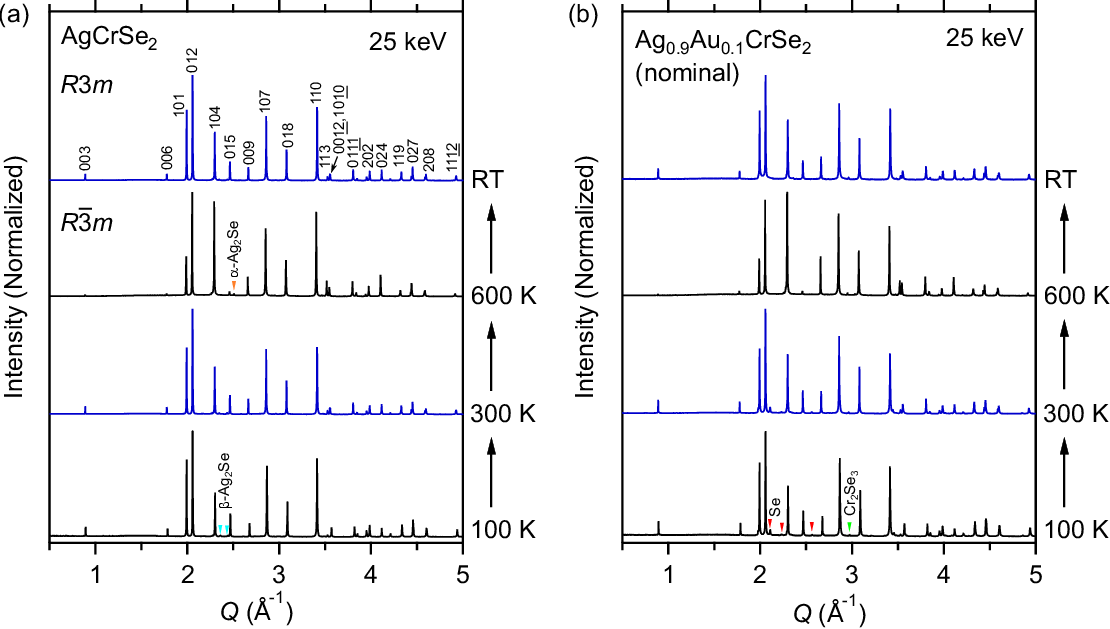}
\caption{\label{sxrd_temp} 
Temperature variation of the powder x-ray diffraction patterns of (a) AgCrSe$_2$ and (b) Ag$_{0.9}$Au$_{0.1}$CrSe$_2$ (nominal composition) obtained at BL02B2 of SPring-8. The experiment was performed during a heating process. After measuring at 600 K, the samples were cooled down to room temperature.
The calibrated wavelength is 0.49630 \AA.
}
\end{center}
\end{figure}

Temperature dependent synchrotron x-ray diffraction experiments were performed on AgCrSe$_2$($x=0$) and Ag$_{0.9}$Au$_{0.1}$CrSe$_2$ (nominal composition) in a temperature range of 100 and 600 K.
Samples were kept at each temperature for 30-60 seconds before scanned.
The results are shown in Figs. \ref{sxrd_temp}(a) and \ref{sxrd_temp}(b).
The top profile in each panel represents the diffraction pattern of the sample cooled down to room temperature after the measurement at 600 K.

First, a small amount of Ag$_2$Se ($\beta$-phase) is observed as a secondary phase in the diffraction pattern of the $x=0$ measured at 100 K. 
On the other hand, the diffraction pattern of the $x=0.1$ sample (nominal composition) measured at 100 K shows secondary phases of Se and Cr$_2$Se$_3$.
Among these, Ag$_2$Se and Se are relatively common impurity phases. 
However, Cr$_2$Se$_3$ is considered to be an impurity phase resulting from the addition of Au beyond its solubility limit.
After heating to 600 K, the Ag$_2$Se ($\beta$-phase) observed in the $x=0$ sample transforms into Ag$_2$Se ($\alpha$-phase). In the diffraction pattern of the $x=0.1$ (nominal) sample at 600 K, Se melts, leaving only Cr$_2$Se$_3$ as the impurity phase.

\begin{figure}[h]
\begin{center}
\includegraphics[width=140mm]{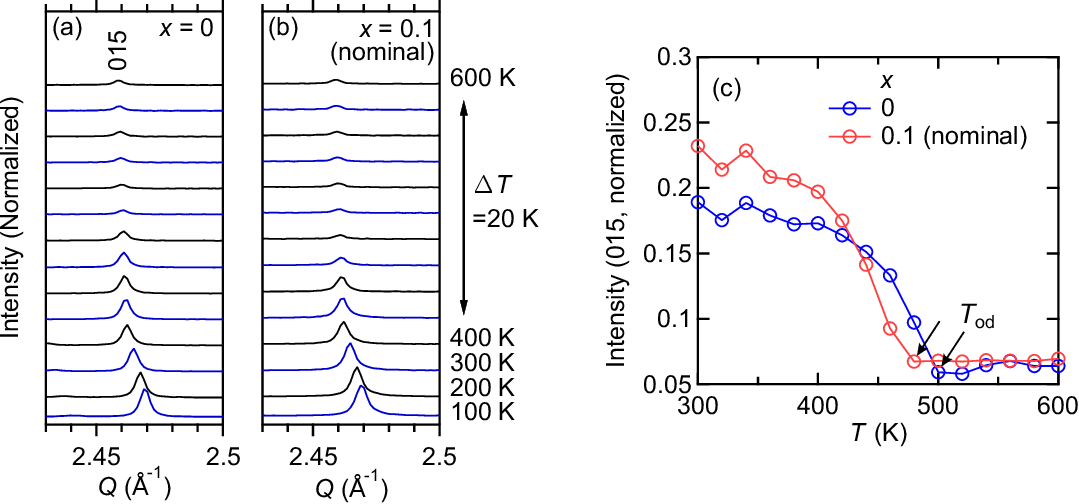}
\caption{\label{015int} 
Intensity of the 015 peak of (a) $x=0$ and (b) $x$(Au) = 0.1 (nominal composition) varying with temperature. The intensity is normalized by the 012 peak intensity for each scan.
(c) The normalized intensity of the 015 peaks plotted as a function of temperature. $T_{\rm od}$ is determined as the temperature at which the intensity curve goes to a flat region, as indicated by arrows. 
}
\end{center}
\end{figure}

The detailed temperature variation of the 015 peak intensity is shown in Figs. \ref{015int}(a) and \ref{015int}(b) for $x=0$ and 0.1(Au, nominal composition), respectively.
The intensity rapidly changes at a temperature both in the $x=0$ and 0.1(nominal composition) samples.
The intensity of the 015 peak is plotted against temperature in Fig. \ref{015int}(c).
Although the Au-addition level exceeds the solid-solubility limit of $x=0.03$ described in the main text, the reduction of the $T_{\rm od}$ is clearly observed in the Ag$_{0.9}$Au$_{0.1}$CrSe$_2$ (nominal composition) sample.

\vspace{5mm}

\clearpage

\section{DSC measurement}

\begin{figure}[h]
\begin{center}
\includegraphics[width=170mm]{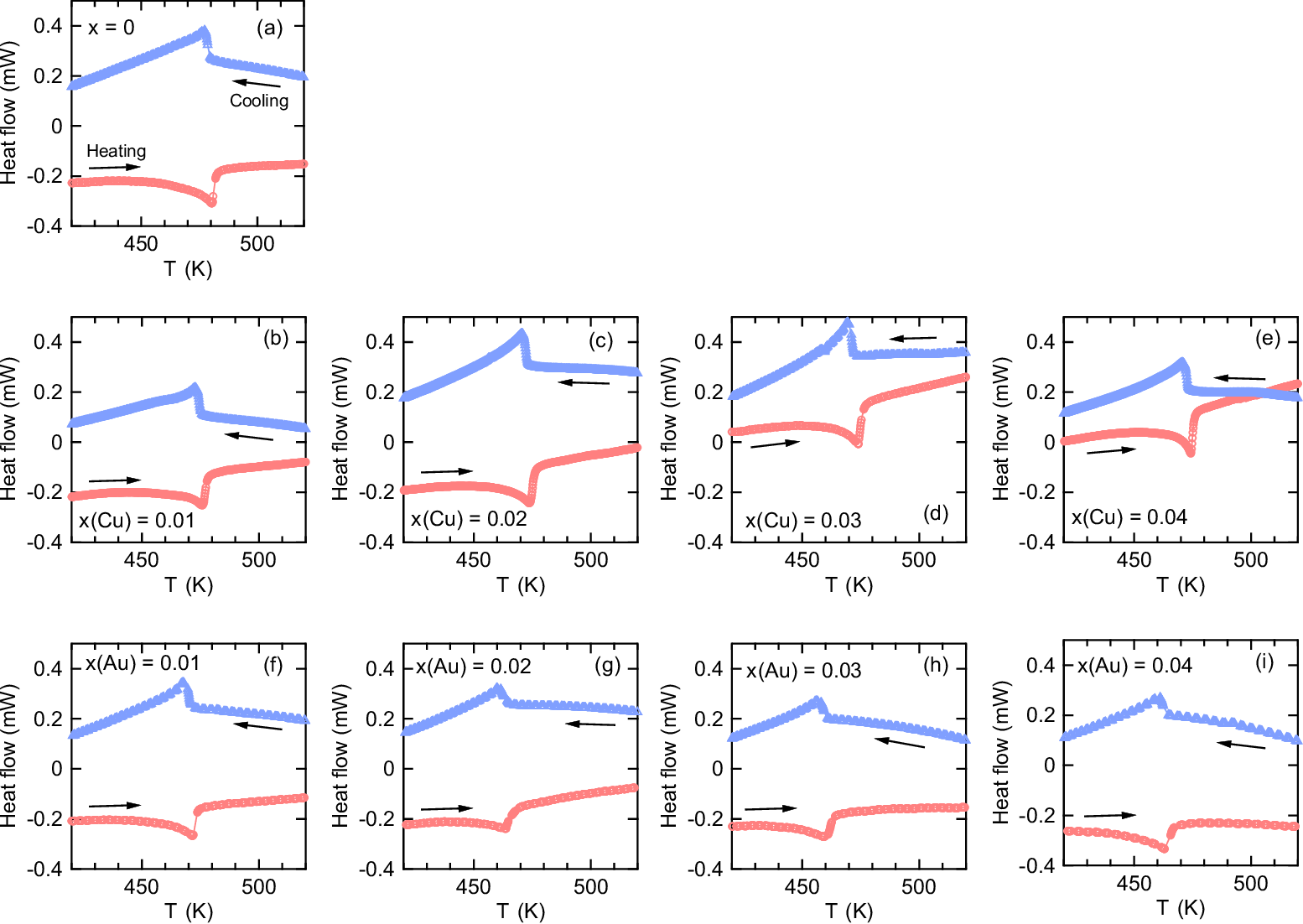}
\caption{\label{DSC_each} 
DSC curves obtained during heating and cooling procedures for (a) $x=0$, (b) $x$ (Cu) = 0.01, (c) $x$ (Cu) = 0.02, (d) $x$ (Cu) = 0.03, (e) $x$ (Cu) = 0.04, (f) $x$ (Au) = 0.01, (g) $x$ (Au) = 0.02, (h) $x$ (Au) = 0.03, and (i) $x$ (Au) = 0.04. The heating curve obtained for $x$ (Cu) = 0.01 shown in (b) is shifted vertically by -0.35 mW for clarity.
}
\end{center}
\end{figure}

	Figs. \ref{DSC_each}(a)-(i) display the DSC curves for $x = 0$, Cu-substituted, and Au-substituted AgCrSe$_2$. 
	The results obtained for both the heating and cooling procedures are shown in each panel. 
	The transition temperature ($T_{\rm od}$) shown in Fig. 3(d) in the main text is defined as the midpoint of the peak positions observed for heating and cooling procedures.

\begin{figure}[h]
\begin{center}
\includegraphics[width=150mm]{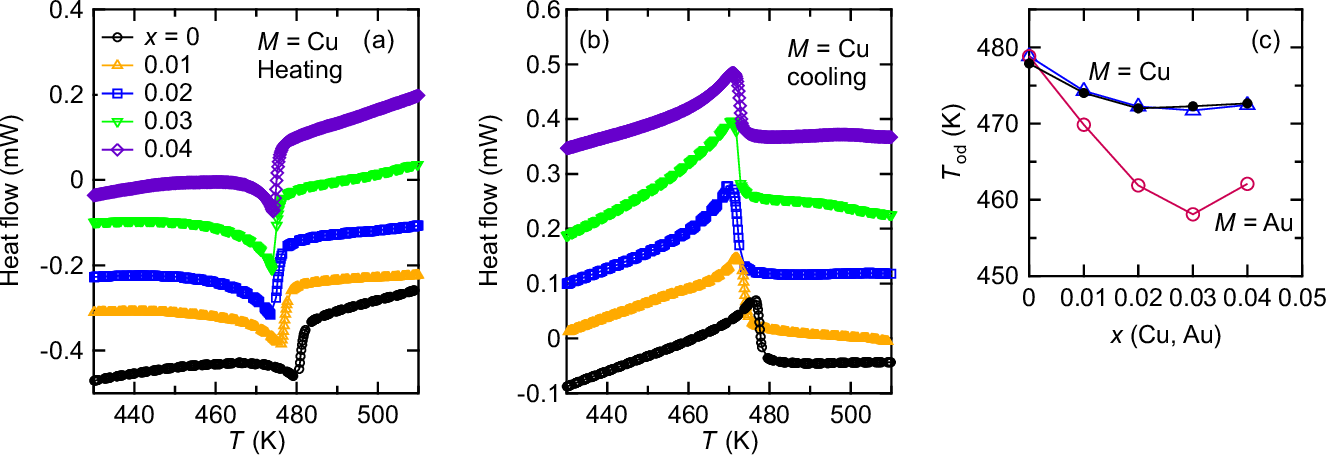}
\caption{\label{DSC_re} 
(a) and (b) show the heating and cooling DSC curves, respectively, measured using a different series of Cu-substituted samples from those described in the main text. (c) $T_{\rm od}$ (filled black circles) determined from the DSC curves shown in (a) and (b). The values of $T_{\rm od}$ shown in Fig. 3(d) in the main text are also plotted for references.
}
\end{center}
\end{figure}

	Figs. \ref{DSC_re}(a) and \ref{DSC_re}(b) show the DSC curves of heating and cooling procedures, respectively, obtained using a different series of Cu-substituted samples from those plotted in Fig. 3(b) in the main text. 
	The $T_{\rm od}$ obtained from these curves are shown in \ref{DSC_re}(c).
	The results reproduce well the trend of $T_{\rm od}$ shown in Fig. 3(d).
	The difference in the $T_{\rm od}$ obtained from two measurements of $x=0$ samples is approximately 1 K, indicating that the $T_{\rm od}$ value has an uncertainty of that magnitude.

\clearpage
\section{Thermal conductivity}

\begin{figure}[h]
\begin{center}
\includegraphics[width=140mm]{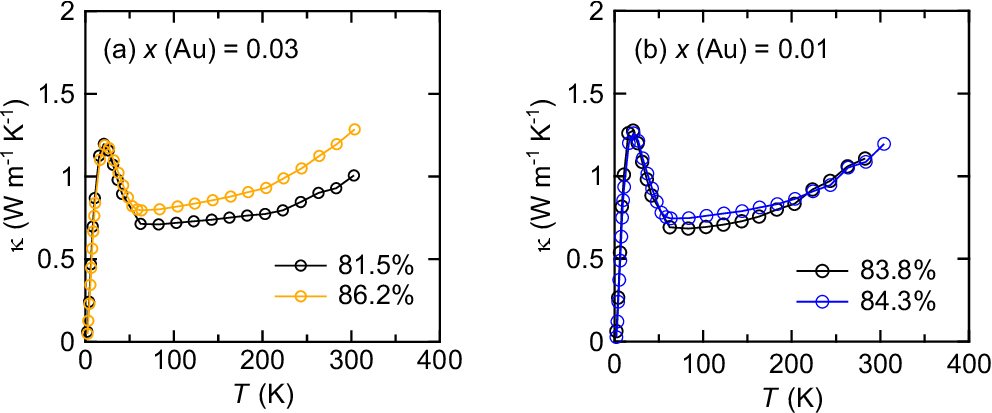}
\caption{\label{kappa1} 
   (a) Thermal conductivity of Ag$_{0.97}$Au$_{0.03}$CrSe$_2$ polycrystals with relative densities of 81.5\% and 86.2\%. 
   (b) Thermal conductivity of Ag$_{0.99}$Au$_{0.01}$CrSe$_2$ polycrystals with relative densities of 83.8\% and 84.3\%. 
}
\end{center}
\end{figure}

\begin{figure}[h]
\begin{center}
\includegraphics[width=130mm]{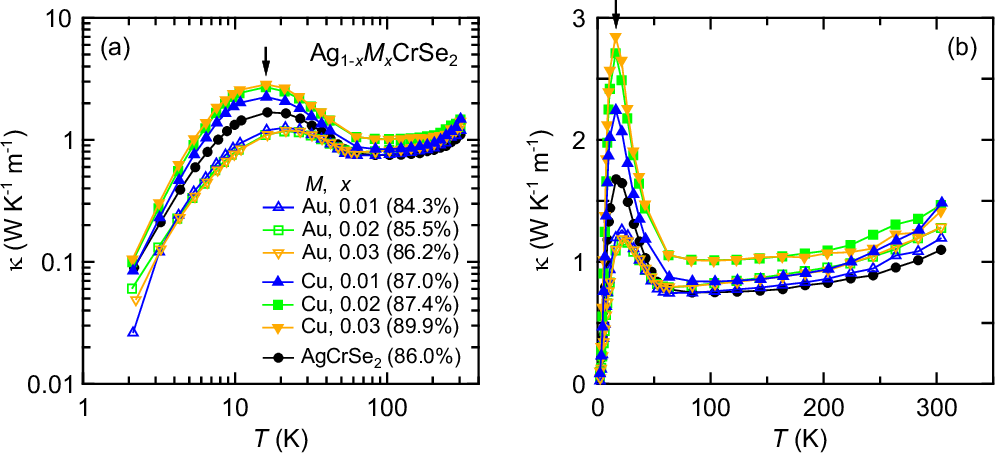}
\caption{\label{kappa2} 
(a) Log-log and (b) linear plots of the thermal conductivity for AgCrSe$_2$ and Ag$_{1-x}M_{x}$CrSe$_2$ ($M$ = Cu and Au) polycrystals. The numbers in the parentheses indicate the relative density of the samples.
}
\end{center}
\end{figure}

Figs. \ref{kappa1}(a) and \ref{kappa1}(b) compares two samples whose relative densities are slightly different to each other. 
As shown in Fig. \ref{kappa1}(a), a small difference in the relative density does not change the thermal conductivity at low temperatures below $\sim$50 K as well as the peak height, although it affects the thermal conductivity at high temperatures.
In Fig. \ref{kappa1}(b), the reproducibility of the samples with almost the same relative densities can be confirmed. 
Figs. \ref{kappa2}(a) and (b) show the log-log and linear plots of the samples for references.

\begin{figure}[h]
\begin{center}
\includegraphics[width=130mm]{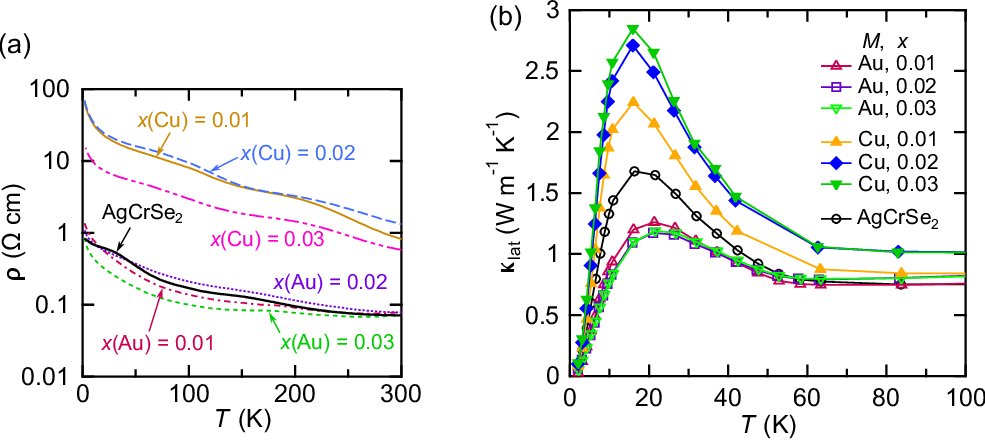}
\caption{\label{kappa_lat} 
(a) Electrical resistivity of AgCrSe$_2$ and Ag$_{1-x}M_{x}$CrSe$_2$ ($M$ = Cu and Au) polycrystals. 
(b) Lattice thermal conductivity ($\kappa_{\rm lat}$) of AgCrSe$_2$ and the Cu- and Au-substituted AgCrSe$_2$. The numbers in the parentheses indicate the relative density of the samples.
}
\end{center}
\end{figure}

	Fig. \ref{kappa_lat}(a) shows the electrical resistivity obtained for $x = 0$ and $M$ = Cu, Au samples. 
	The values are used for evaluating the electronic contribution of thermal conductivity.
	The lattice thermal conductivity is obtained by subtracting $\kappa_{\rm el}$ from $\kappa_{\rm total}$ and is shown in Fig. \ref{kappa_lat}(b).

\vspace{5mm}

\clearpage
\section{Structure refinement for the synchrotron XRD patterns}

\begin{figure}[h]
\begin{center}
\includegraphics[width=120mm]{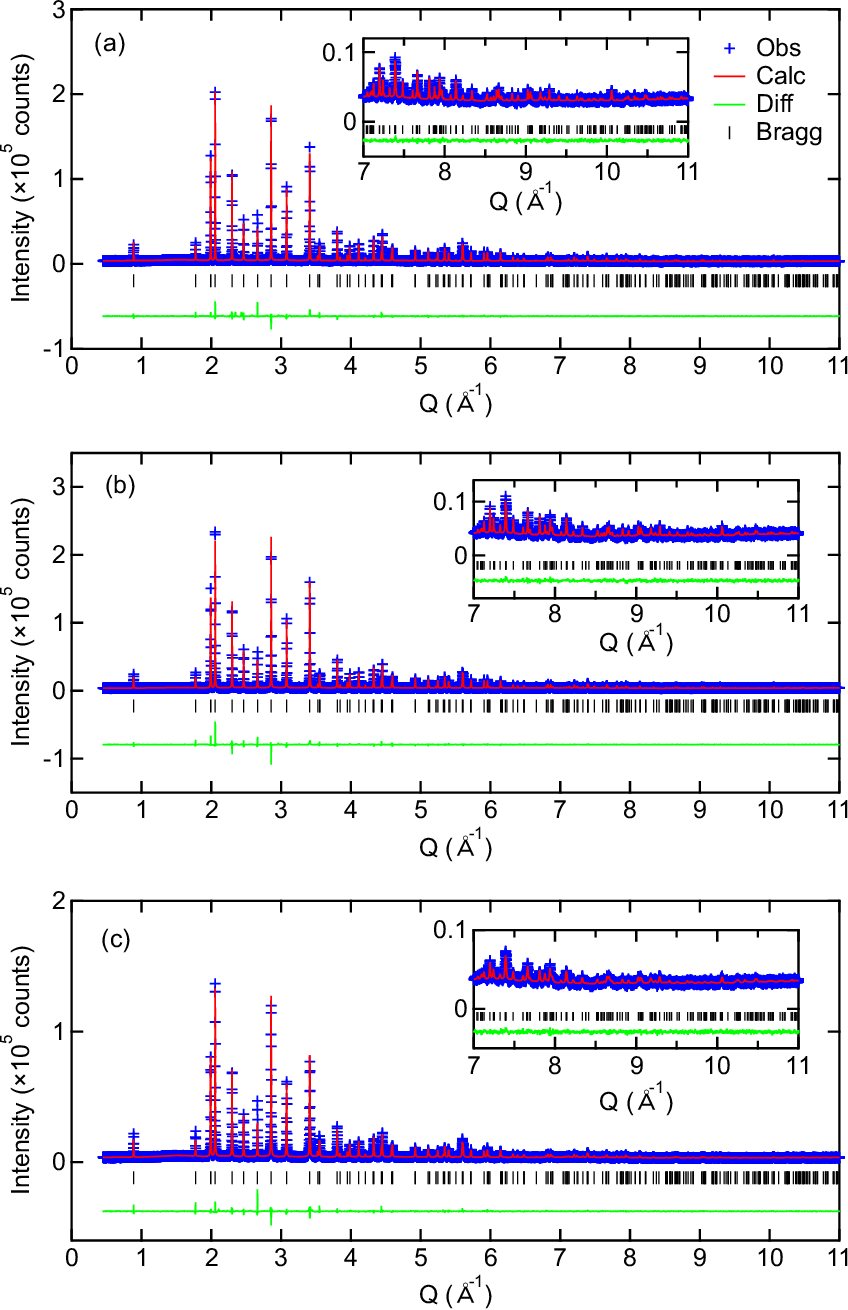}
\caption{\label{sXRD} 
   (a) Synchrotron x-ray diffraction patterns and fitted profiles of (a) AgCrSe$_2$, (b)  Ag$_{0.97}$Cu$_{0.03}$CrSe$_2$, and (c) Ag$_{0.97}$Au$_{0.03}$CrSe$_2$. The incidence x-ray energy is 24 keV ($\lambda$ = 0.51689 \AA, calibrated).
}
\end{center}
\end{figure}

Powder x-ray diffraction was carried out at the BL19B2 beamline (SPring-8) at room temperature.
The x-ray energy was set at 24 keV. 
Fine powder was filled into a fused quartz capillary with a diameter of 0.2 mm for the measurements.
Obtained diffraction data were fitted using FullProf software.
Figs. \ref{sXRD}(a)--(c) display the experimental diffraction patterns and the fitted profiles of AgCrSe$_2$ and Ag$_{0.97}M_{0.03}$CrSe$_2$ ($M$ = Cu and Au).

